\documentclass[a4paper]{jpconf}

\usepackage{graphicx}
\usepackage{amssymb}
\usepackage{epsfig}
\newcommand{\postscript}[2]{\setlength{\epsfxsize}{#2\hsize}
   \centerline{\epsfbox{#1}}}

\newcommand{\beq}[1]{\begin{equation}\label{#1}}
\newcommand{\eeq}{\end{equation}}
\newcommand{\bea}[1]{\begin{eqnarray}\label{#1}}
\newcommand{\eea}{\end{eqnarray}}
\newcommand{\ba}{\begin{array}}
\newcommand{\ea}{\end{array}}

\newcommand{\be}{\begin{equation}}
\newcommand{\ee}{\end{equation}}

\begin{document}

\title{Large Scale Anisotropy of Cosmic Rays and Directional Neutrino Signals from Galactic Sources}
\author{
Luis A. Anchordoqui$^{1}$,
Haim Goldberg$^2$
Angela V. Olinto$^{3,4}$,\\
Thomas C. Paul$^{1,2}$,
Brian J. Vlcek$^1$
and
Thomas J. Weiler$^{5}$
}

\address{
$^1$ Department of Physics, University of Wisconsin-Milwaukee, Milwaukee,  WI, 53201 USA\\
$^2$ Department of Physics, Northeastern University, Boston, MA 02115, USA\\
$^3$ Department of Astronomy and Astrophysics,  Enrico Fermi Institute,  Chicago, Il 60637, USA\\
$^4$ Kavli Institute for Cosmological Physics, University of Chicago, Chicago, Il 60637, USA\\
$^5$ Department of Physics \& Astronomy, Vanderbilt University, Nashville, TN, 37235 USA

}

\def\myead#1{\vspace*{5pt}\address{Speaker's E-mail: \mailto{#1}}}
\myead{luis.anchordoqui@gmail.com}

\begin{abstract}
  We investigate the neutrino $\leftrightharpoons$ cosmic ray
  connection for sources in the Galaxy in terms of two observables: the shape of the
  energy spectrum and the distribution of arrival directions. We also study
  the associated gamma ray emission from these sources.
\end{abstract}

Above about 10~GeV, cosmic ray (CR) energy spectrum observed at Earth
falls roughly as a power law, with flux decreasing almost three orders of
magnitude per energy decade until eventually it exhibits a strong
suppression near 60~EeV~\cite{Abbasi:2007sv}. Closer examination
reveals several other spectral features. A steepening of the spectrum
between $J(E) \propto E^{-2.67\pm 0.07}$ and $E^{-3.07 \pm 0.11}$ has been
labeled the ``knee'' and is prominent at $E_{\rm knee} \approx 3~{\rm
  PeV}$~\cite{Hoerandel:2002yg}.  A less prominent ``second knee'',
corresponding to additional softening of the spectrum, $J (E) \propto E^{-3.52 \pm
  0.19}$, appears above 0.3~EeV~\cite{AbuZayyad:2000ay}. At $E_{\rm
  ankle} \approx 3~{\rm EeV}$ a pronounced hardening of the spectrum
appears, the so-called ``ankle'' feature~\cite{Bird:1993yi}.

The Earth lies near the edge of the Galaxy, so if CR sources are
concentrated in the Galactic disk one would expect to
observe a dipole anisotropy in the flux, with a relative
excess in the direction of the Galactic center.  Hints of this
anisotropy have in fact emerged. The IceCube Collaboration
has reported~\cite{Abbasi:2011zka} an excess of events in the 400~TeV
sky-map ($29^\circ$  smoothing) at right ascension $\equiv
\alpha = 256.6^\circ$ and declination $\equiv \delta = -25.9^\circ$,
with a pre-trial significance of $5.3\sigma$. The data also show a
deficit at $\alpha = 73.1^\circ$ and $\delta = -25.3^\circ$, with a
pre-trial significance value of $8.6\sigma$ (optimized smoothing of
$21^\circ$).  After correcting for
the trials, only the deficit remains significant beyond the $5\sigma$
level, with a post-trial significance of $6.3\sigma$.

The right ascension distribution of the flux
can be characterized by the amplitudes and phases of its Fourier
expansion~\cite{Linsley:1975kp}
\begin{equation}
J (\alpha) = J_0 [1 + {\cal A} \cos (\alpha - \phi)  + {\cal A}' \cos (2 (\alpha - \phi')) + \dots \,] \,,
\end{equation}
where $J_0$ is the monopole intensity, ${\cal A}$ and ${\cal A}'$ are
the first and second harmonic amplitudes and $\phi$ and $\phi'$  the
associated phases. The right ascension harmonic analysis cannot reveal the component of
the anisotropy vector $\vec \delta$ along the Earth rotation axis,
$\delta_\parallel = \vec \delta \, \sin \lambda$, where $\lambda$ is
the latitude of the direction where the flux is maximum. The first
harmonic amplitude ${\cal A}$ is related to the component of
anisotropy in the equatorial plane, $\delta_\perp = \vec \delta \,
\cos \lambda,$ via ${\cal A} \simeq \delta_\perp \
\overline{c_\delta}$, where $\overline{c_\delta}$ is the mean value of
the cosine of the event declinations~\cite{Aublin:2005nv}.
The first harmonic amplitude reported by the IceCube Collaboration is
${\cal A} = (3.7 \pm 0.7_{\rm stat.} \pm 0.7_{\rm stat}) \times
10^{-4}$ and the associated phase is $\phi = 239 \pm 10.6_{\rm stat.}
\pm 10.8_{\rm syst.}$. These results are in agreement with previous
observations by EAS-TOP Collaboration which, for 370~TeV, reported
${\cal A} = (6.4 \pm 2.5) \times 10^{-4}$ with $\phi = 204.0 \pm
22.5$~\cite{Aglietta:2009mu}.

Recently, the Pierre Auger Collaboration reported an analysis of the
first harmonic modulation in the right ascension distribution of the
events recorded from 1/1/2004 to 12/31/2010 with the surface detector
(SD) 
array, and from 9/12/2007 to 4/11/2011 with the infill
array~\cite{ThePierreAuger:2013eja}. This analysis takes advantage of
the wide range of energy (10 PeV -- 100 EeV) that the Pierre Auger
Observatory is able to scan thanks to the infill array. While no clear
evidence for anisotropy has been found, yet it is interesting to note
that in the range above 1~EeV, 3 out of the 4 energy bins are above
the 99\% C.L. expectation from isotropy, {\it i.e.} only one percent
of isotropic samples would show equal or larger amplitudes. The phase
evolution in this wide energy range has an interesting behavior, with
a smooth transition from a common phase of $\phi = 270^\circ$ in the
bins below $1$~EeV to a phase $\phi = 90^\circ$ above 5~EeV.  The
phase at lower energies is compatible with the right ascension of the
Galactic center, $\alpha_{\rm GC} \simeq 268.4^\circ$. To test the
hypothesis that the phase is undergoing a smooth transition, the
Pierre Auger Collaboration began to independently analyze data
obtained after April 2011. After 18 months the new and independent
data set is showing a similar trend~\cite{ThePierreAuger:2013eja}.
Another 18 months of data collection to reach an aperture of
21,000\,km$^2$\,sr with the independent data set is needed before the
trend can be confirmed.  It is interesting to note that despite the
possible hints for cosmic ray anisotropy discussed above, any such
anisotropy would be remarkably small (at the \% level). The existing
limits on the equatorial amplitude $\delta_\perp$ and the reported
phases by the various experiments are shown in Fig.~{\ref{fig:ani}.

To quantify the spectral features characteristic of  Galactic CR models we
adopt the ``leaky box'' picture, in which CRs propagate freely in the
Galaxy, contained by the magnetic field but with some probability to
escape which is constant in time. The local energy density is given by
\begin{equation}
n_{\rm CR} (E) \equiv \frac{4\pi}{c} J (E) \, \approx Q(E) \
\tau(E/Z),
\label{leaky}
\end{equation}
where $Q(E) \propto E^{-\alpha}$ is the generation rate of primary CRs
and $\tau (E/Z) \propto E^{-\delta}$ is the rigidity-dependent
confinement time (for details, see
e.g.~\cite{Anchordoqui:2013dnh}). Fits to the energy dependence of
secondary to primary ratios yield
$\delta=0.6$~\cite{Swordy:1993dz}. For a source index $\alpha \simeq
2.07$, which is close to the prediction of Fermi shock acceleration,
inclusion of propagation effects reproduces the observed
spectrum. However, $\delta = 0.6$ results in an excessively large
anisotropy which is inconsistent with the upper limits shown in
Fig.~\ref{fig:ani}~\cite{Blasi:2011fm}. Consistency with anisotropy can be
achieved by adopting a Kolmogorov index, $\delta =
1/3$~\cite{Biermann:1995qy,Candia:2003dk}. The apparent conflict with
the secondary to primary composition analyses can be alleviated
through small variations of the energy dependence of the spallation
cross sections, or variation in the matter distribution in the
Galaxy~\cite{Biermann:1995qy}.  This hypothesis implies a steeper
source spectrum, $\alpha \simeq 2.34$.

\begin{figure}[tbp]
\begin{minipage}[t]{0.49\textwidth}
\postscript{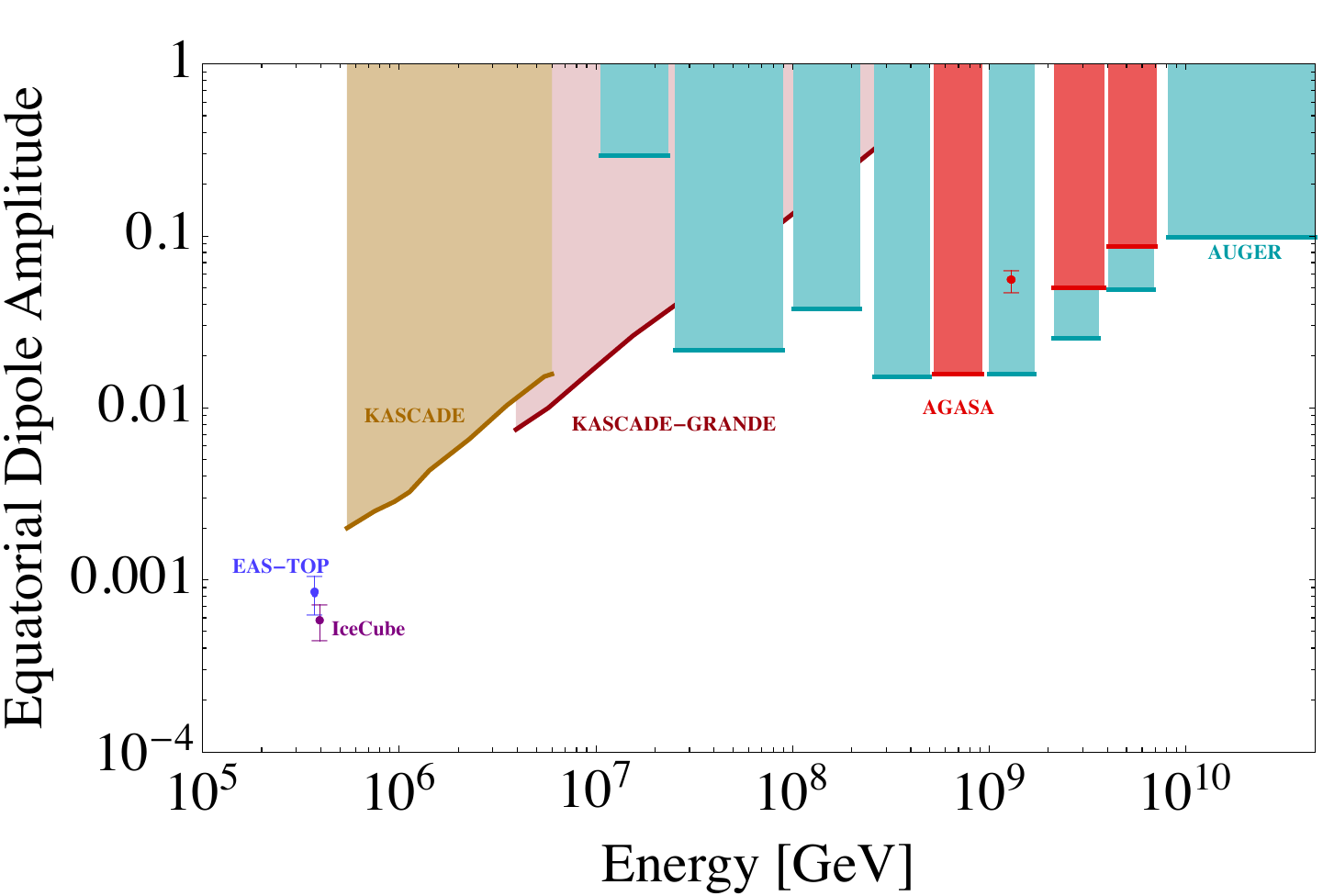}{0.99}
\end{minipage}
\hfill
\begin{minipage}[t]{0.49\textwidth}
\postscript{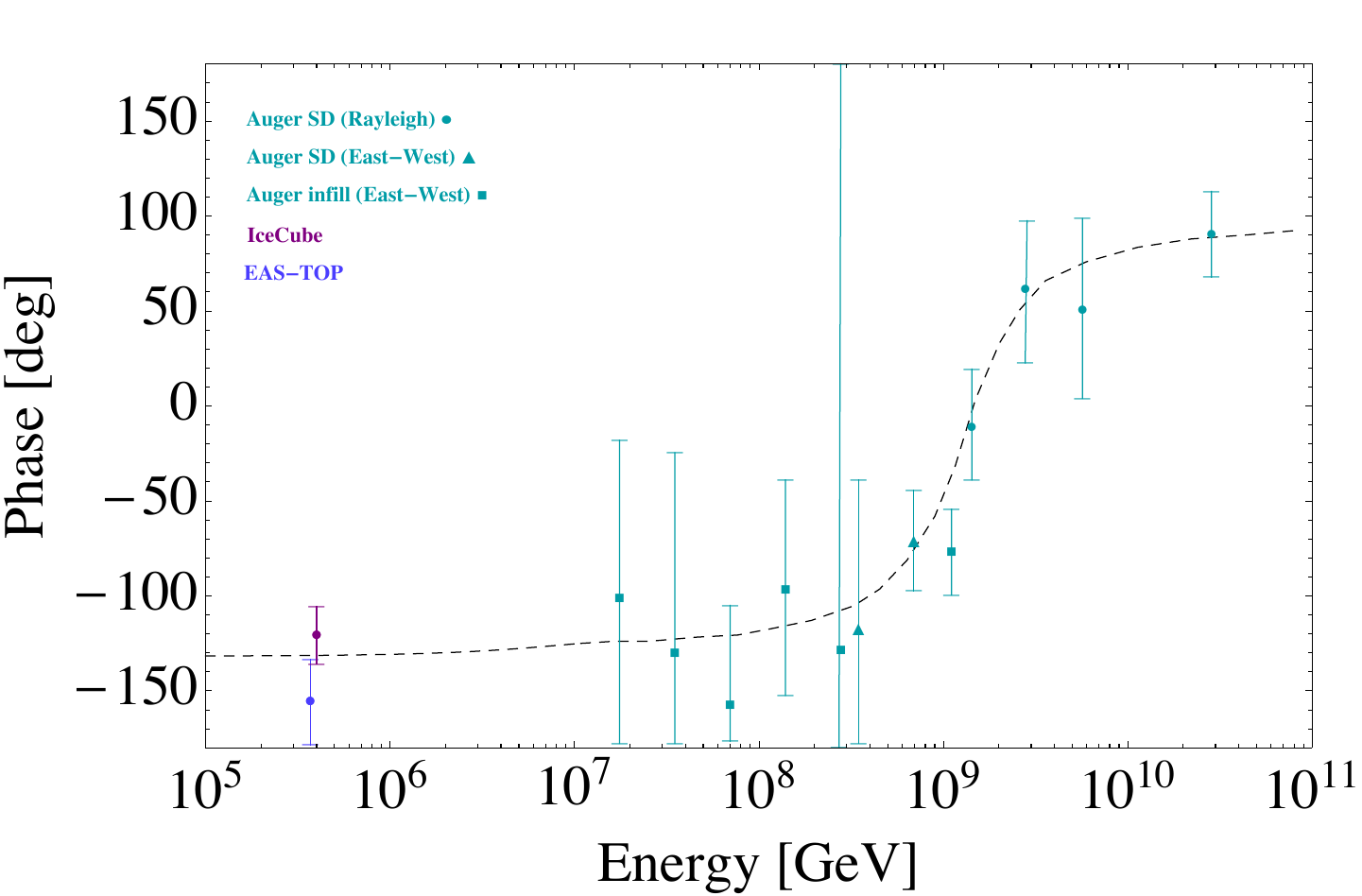}{0.99}
\end{minipage}
\caption{{\bf Left:} Measurements of the first harmonic amplitude
  (corrected by the mean value of the cosine of the event
  declinations) by IceCube~\cite{Abbasi:2011zka},
  EAS-TOP~\cite{Aglietta:2009mu}, and AGASA~\cite{Hayashida:1998qb}.  The shaded regions are
  excluded by null results of searches by KASCADE (@ 95\%
  C.L.)~\cite{Antoni:2003jm} and KASCADE-Grande (@ 95\%
  C.L.)~\cite{Stumpert:2008zz}, Auger (@ 99\%
  C.L.)~\cite{ThePierreAuger:2013eja}, and AGASA collaborations~\cite{Hayashida:1998qb}.  {\bf
    Right:} Phase of the first harmonic as a function of energy as
  reported by the IceCube~\cite{Abbasi:2011zka}, the
  EAS-TOP~\cite{Aglietta:2009mu}, and the Pierre
  Auger~\cite{ThePierreAuger:2013eja} collaborations. The phase shows
  a smooth evolution from the Galactic center towards the Galactic
  anti-center region.}
\label{fig:ani}
\end{figure}

It is helpful to envision the CR engines as machines where protons are
accelerated and (possibly) permanently confined by the magnetic fields
of the acceleration region. The production of neutrons and pions and
subsequent decay produces neutrinos, $\gamma$-rays, and CRs. If the
CR source also produces high energy neutrinos, then pion
production must be the principal agent for the high energy cutoff on
the proton spectrum.  Conversely, since the protons must undergo
sufficient acceleration, inelastic pion production needs to be small
below the cutoff energy; consequently, the plasma must be optically
thin. Since the interaction time for protons is greatly increased over
that of neutrons due to magnetic confinement, the neutrons escape
before interacting, and on decay give rise to the observed CR
flux. The foregoing can be summarized as three conditions on the
characteristic nucleon interaction time scale $\tau_{\rm int}$; the
neutron decay lifetime $\tau_n$; the characteristic cycle time of
confinement $\tau_{\rm cycle}$; and the total proton confinement time
$\tau_{\rm conf}$: $(i)\; \tau_{\rm int}\gg \tau_{\rm cycle}$; $(ii)\;
\tau_n > \tau_{\rm cycle}$; $(iii)\; \tau_{\rm int}\ll \tau_{\rm
  conf}$. The first condition ensures that the protons attain
sufficient energy.  Conditions $(i)$ and $(ii)$ allow the neutrons to
escape the source before decaying. Condition $(iii)$ permits
sufficient interaction to produce neutrons and neutrinos. These three
conditions together define an optically thin source~\cite{Ahlers:2005sn}.

If  CR sources in the Galaxy are optically thin, then one would
expect a flux of neutrinos with a spectrum $\propto E^{-2.34}$. If
the maximum attainable energy of cosmic rays in the Galaxy is $E_{\rm
  max} \sim 1 - 3~{\rm EeV}$, by Hillas criterion~\cite{Hillas:1985is} we would expect protons to be accelerated to
$E_{\rm max}^p = E_{\rm max}/26 \sim 40~ {\rm PeV}$ and therefore secondary neutrinos to be
be produced with energies of up to 2 to 3~PeV.

Quite recently the IceCube Collaboration has reported a preliminary
observation of 26 neutrino candidates~\cite{Aartsen:2013jdh}, in addition to but at lower energies than
the two $\sim 1$~PeV neutrinos reported earlier in 2013~\cite{Aartsen:2013bka}. Taken together, these
28 events constitute a 4.1$\sigma$ excess compared to expectations from
background. Interpretation of these results, however, does not appear to be
entirely straightforward.  For instance, if one makes the common assumption of
an unbroken $E_\nu^{-2}$ neutrino energy spectrum, then one expects to observe about
8-9 events with higher energies than the two highest energy events observed thus
far. Recently, we have
investigated the compatibility between the IceCube observations and
the hypothesis of an unbroken power-law spectrum arising from
optically thin Galactic neutrino sources, finding that a  cosmic neutrino
flux per flavor,  averaged over all three flavors,
\begin{equation}
\frac{dF_\nu}{d\Omega dA dt dE_\nu} = 6.62 \times 10^{-7} \
\left(\frac{E_\nu}{1~{\rm GeV}} \right)^{-2.3}~({\rm GeV}\cdot{\rm cm}^{2}\cdot{\rm s}\cdot{\rm sr})^{-1}\,,
\label{eqn:flux}
\end{equation}
is consistent with the data reported thus
far~\cite{Anchordoqui:2013qsi}.

The assumption underlying the leaky box model is that the energy
density in CRs observed locally is typical of other regions of the
Galactic disk. If so, the total power required to maintain the cosmic
radiation in equilibrium can be obtained by integrating the generation
rate of primary CRs over energy and space. Using (\ref{leaky}), we
obtain
\begin{equation}
\frac{d\epsilon_{\rm CR}}{dt} = \int d^3x \int Q(E) \: dE = V_G \frac{4\pi}{c}
\int \frac{J(E)}{ \tau(E/Z)}  dE \, ,
\end{equation}
where $V_G \sim 10^{67}~{\rm cm}^3$ is the Galactic disk
volume~\cite{Gaisser:2005tu}. For $E_{\rm knee} < E < E_{\rm ankle}$,
we conservatively assume that the trapping time in the Galaxy scales
with energy as $\tau = 2 \times 10^{7} (E_{{\rm GeV}}/Z)^{-1/3}~{\rm
  yr}$~\cite{Gaisser:2006xs}.  In this
case the power budget required to fill in the spectrum from the knee
to the ankle is found to be $d\epsilon_{\rm CR}/dt \simeq 2 \times
10^{39}~{\rm erg/s}$~\cite{Gaisser:2006xs}. 
From this, we deduced elsewhere ~\cite{Anchordoqui:2013qsi}
the energy transfer fraction from the parent protons to the pions which
ultimately produce the observed neutrinos, demonstrating that $pp$
collisions are more likely to produce the neutrino flux than are $p
\gamma$ collisions, and indeed that the power budget is enough to reproduce the
neutrino flux in (\ref{eqn:flux}).

It is interesting to employ existing limits on high energy photons to
check the plausibility of our hypothesis that the IceCube excess is of
Galactic origin.  $\gamma$ rays are produced by $\pi^0$ decays at the
same optically thin sources where neutrinos are produced by
$\pi^{\pm}$ decay. As described in~\cite{Anchordoqui:2013dnh}, one
can predict a differential $\gamma$ ray flux based the best-fit single
power law $\nu$ flux discussed in this paper, and compare to
measurements.  The CASA-MIA 90\% C.L. upper limits on the integral
diffuse $\gamma$ ray flux, $I_\gamma$ for energy bins,
\begin{equation}
\frac{E_\gamma^{\rm min}}{{\rm GeV}} = 3.30 \times 10^5,\ 7.75 \times 10^5,\ 2.450 \times 10^6 \,,
\label{tres}
\end{equation}
are
\begin{equation}
\frac{I_\gamma}{{\rm cm}^{-2} \ {\rm s}^{-1} \ {\rm  sr}^{-1}}  <  1.0 \times 10^{-13}, \, 2.6 \times 10^{-14},  2.1 \times 10^{-15}\,,
\label{cuatro}
\end{equation}
respectively~\cite{Chantell:1997gs}.
A more rigorous comparison would involve measurements on the diffuse $\gamma$
fay flux within about $15^\circ$ of the Galactic plane. The CASA-MIA Collaboration has
in fact studied $\gamma$ ray emission from the direction of the Galactic plane,
reporting the flux limits as a fraction of the CR flux~\cite{Borione:1997fy}
rather than an integral bound.  Comparing the relative fractions from the
all-sky analysis to the Galactic plane analysis indicates that constraining the
observation to the Galactic plane region does indeed lead to tighter
constraints:
\begin{equation}
\frac{I_\gamma}{{\rm cm}^{-2} \ {\rm s}^{-1} \ {\rm  sr}^{-1}}  \lesssim
5.0 \times 10^{-14}, \, 1.3 \times 10^{-14}, \  2.1 \times 10^{-15}\,,
\label{4cinco}
\end{equation}
at the 90\% C.L.
Under the simplifying assumption that there is no
photon absorption, the integral photon fluxes we predict based on our single power law hypothesis
(in units of photons ${\rm cm}^{-2} \ {\rm s}^{-1} \ {\rm sr}^{-1}$), above the
energies specified in (\ref{tres}), are
\begin{equation}
\int_{E_\gamma^{\rm min}} \frac{dF_\gamma}{d\Omega dA dt dE_\gamma} dE_\gamma  =  4.2 \times 10^{-14}, \ 1.4 \times 10^{-14}, \  3.1 \times 10^{-15} \, .
\label{cinco}
\end{equation}
For the first two energy bins, the predicted fluxes saturate the
90\%~C.L. measurements of CASA-MIA, while the last bin slightly exceeds the 90\%
C.L. bound. This does not, however, imply that the Galactic origin hypothesis is
ruled out at 90\% C.L.  First of all, one must keep in mind that sources which
are optically thin up to $E_\gamma \sim 100$~TeV may not be optically thin at
higher energies, suggesting that the importance of photon bounds in
establishing the origin of the IceCube excess should be considered with some
caution.  Even if we ignore this caveat, we still do not know the maximum
neutrino energy reached at acceleration sites, so the maximum photon energy is
likewise unknown. In addition, absorption becomes important in the energy regime
covered by the last bin, as mean free path of PeV photons in the CMB is about 10~kpc.

Note that $R_{\rm G} \sim 10~{\rm kpc}$,  leading to an interesting
signature: Photons coming from ``our half'' of the Galaxy will be largely
unattenuated, while those from the farther half will be significantly
attenuated.  Since both photons and neutrinos point back to the sources,
coordinated comparisons of neutrino and photon data will facilitate a completely
new exploration of the highest-energy Galactic sources.  As described
in~\cite{Anchordoqui:2013dnh}, taking into account absorption of the photon flux
for $E_\gamma^{\rm min} > 1$~PeV leads to about a 12\% reduction in the
predicted photon flux.  Furthermore, varying the photon maximum energy cutoff of
Eq.~(\ref{cinco}) to,
\begin{equation}
\frac{E_\gamma^{\rm max}}{\rm PeV} = 6, \ 7, \ 8  \,,
\end{equation}
we obtain
\begin{equation}
\int_{E_\gamma^{\rm min} }^{E_\gamma^{\rm max}}
\frac{dF_\gamma}{d\Omega dA dt dE_\gamma} dE_\gamma   =   2.1 \times
10^{-15},\ 2.3 \times 10^{-15}, \  2.4 \times 10^{-15} \, .
\end{equation}
From the discussion above, we can see there are several ways to comply with the
CASA-MIA bound.  For instance, $E_\gamma^{\rm max} = 6~{\rm PeV}$ is already
consistent with the measured bound, even without absorption.  For higher
energies, absorption provides enough reduction of the photon flux to retain
consistency with measurements.  It is also worth noting that the comparison
discussed here is based on experimental bounds on the all-sky $\gamma$
ray flux.

Only the IceCube Collaboration has thus far reported constraints on
$\gamma$ ray emission
between 1
and 10 PeV from the direction of the Galactic plane.  Bounds from the IceCube 40 string
configuration~\cite{Aartsen:2012gka}, are not restrictive enough to challenge
the Galactic origin hypothesis.  However, within 5 years of data taking with the
complete IceCube configuration of 86 strings, enough statistics will be gathered
to elucidate the $\nu - \gamma$ ray connections.

Finally we comment on the consistency between the arrival direction
distribution of the IceCube excess and the hypothesis that the sources
are nearby.  Fourteen of the 26 reported neutrino events arrive from
within about $15^{\circ}$ of the Galactic plane, including one of the
two highest energy events, which coincides with the Galactic center
(within errors)~\cite{Razzaque:2013uoa} .  The highest energy event is outside of this angular
window, but (as noted in~\cite{Ahlers:2013xia}) does correspond with a
possible hotspot in the IceCube photon search~\cite{Aartsen:2012gka}.
This could reflect emission of neutrinos and $\gamma$ rays from a
common, nearby source, as $\gamma$ rays do not survive propagation
further than $\sim 10$~kpc. The recently discovered large reservoir of
ionized gas extending over a large region around the Milky
Way~\cite{Gupta:2012rh} could provide the target material required for
neutrino production outside the Galactic disk in models in which
proton diffusion extends to the Galactic 
halo~\cite{Jones:2000qd,Taylor:2014hya}. However, given the current statistics and
the insufficient understanding of the atmospheric (in particular the
prompt neutrino~\cite{Lipari:2013taa}) background, the arrival
direction distribution neither favors nor disfavors a Galactic
origin~\cite{Ahlers:2013xia,Neronov:2013lza}.  More data are required
to settle the issue.

\ack This work was supported by the US NSF grant numbers: CAREER
PHY1053663 (LAA); PHY-0757959 (HG); PHY-1205854 (TCP); PHY-1068696,
PHY- 1125897 (AVO); US DoE grant DE-FG05-85ER40226 (TJW); NASA grant numbers:
NNX13AH52G (LAA, TCP), 11-APRA- 0066 (AVO); the Simons Foundation
(grant \#306329) (TJW); and the UWM RGI (BJV).

\end{document}